\documentclass[twocolumn,showpacs,floats,floatfix,superscriptaddress,aps,prl]{revtex4-1}
\usepackage{amsfonts,amssymb,amsmath}
\usepackage{color,calc}
\usepackage[dvips]{graphicx}
\usepackage{bm}
\usepackage[normalem]{ulem}

\def\i{\,\text{i}}
\def\e{\,\text{e}}
\def\i{i}
\def\e{e}

\def\to{\rightarrow}

\def\U{\mathbf{U}}

\def\to{\rightarrow}

\def\i{\text{i}}
\def\f{\text{f}}

\newcommand{\pryso}{$\text{Pr}^{3+}\text{:}\text{Y}_2\text{SiO}_5\:$}

\begin{document}

\author{Genko T. Genov}
\affiliation{Institut f{\"u}r Angewandte Physik, Technische Universit{\"a}t Darmstadt, Hochschulstr. 6, 64289 Darmstadt, Germany}
\author{Daniel Schraft}
\affiliation{Institut f{\"u}r Angewandte Physik, Technische Universit{\"a}t Darmstadt, Hochschulstr. 6, 64289 Darmstadt, Germany}
\author{Nikolay V. Vitanov}
\affiliation{Department of Physics, St. Kliment Ohridski University of Sofia, 5 James Bourchier blvd, 1164 Sofia, Bulgaria}
\author{Thomas Halfmann}
\affiliation{Institut f{\"u}r Angewandte Physik, Technische Universit{\"a}t Darmstadt, Hochschulstr. 6, 64289 Darmstadt, Germany}

\title{Arbitrarily Accurate Pulse Sequences for Robust Dynamical Decoupling}

\date{\today}

\begin{abstract}
We introduce universally robust sequences for dynamical decoupling, which simultaneously compensate pulse imperfections and the detrimental effect of a dephasing environment to an arbitrary order, work with any pulse shape, and improve performance for any initial condition.
Moreover, the number of pulses in a sequence grows only linearly with the order of error compensation. Our sequences outperform the state-of-the-art robust sequences for dynamical decoupling. Beyond the theoretical proposal, we also present convincing experimental data for dynamical decoupling of atomic coherences in a solid-state optical memory.
\end{abstract}

\maketitle


\emph{\textbf{Introduction.}}--
Quantum technologies are increasingly important nowadays for a multitude of applications in
sensing, processing, and communication of information. Nevertheless, protection of quantum systems from unwanted interactions with the environment remains a major challenge. Dynamical decoupling (DD) is a widely used approach that aims to do this by nullifying the average effect of the unwanted qubit-environment coupling through the application of appropriate sequences of pulses
\cite{Viola09PRL,Viola13NatComm,Khodjasteh05PRL,Khodjasteh07PRA}.

Most DD schemes focus on dephasing processes because they have maximum contribution to information loss in many systems, e.g., in nuclear magnetic resonance and quantum information \cite{Viola16NJP,RDD_review12Suter}. Then, the major limitation to DD are pulse imperfections whose impact often exceeds the effect of the perturbations from the environment \cite{RDD_review12Suter,Suter_DD,Suter_DD_light_storage}. Some sequences, e.g., the widely used Carr-Purcell-Meiboom-Gill (CPMG) sequence, work efficiently for specific quantum states only \cite{RDD_review12Suter,CPMG_papers}. Robust sequences for any state with limited error compensation have been demonstrated experimentally, e.g., XY4 (PDD), Knill DD (KDD) \cite{RDD_review12Suter}. Composite pulses, designed for static errors, were also recently shown to be robust to time-dependent non-Markovian noise up to a noise frequency threshold \cite{Viola14PRA}. A common feature of most robust DD sequences so far is pulse error compensation in one or two parameters only (flip angle error, detuning). High fidelity error compensation has been proposed, e.g., by nesting of sequences, but only at the price of a very fast growth in the number of pulses \cite{RDD_review12Suter}.

In this Letter, we describe a general theoretical procedure to derive universally robust (UR) DD sequences that compensate pulse imperfections in any experimental parameter (e.g., variations of pulse shapes or intensities), and the effect of a slowly changing environment to an \emph{arbitrary order} in the permitted error. We note that the term universal is applied for pulse errors. The UR sequences work at high efficiency for any initial condition. The number of pulses for higher order error compensation grows only linearly with the order of the residual error. The concept works for arbitrary pulse shapes. Our only assumptions are identical pulses in a sequence and a correlation time of the environment that is longer than the sequence duration -- in order to maintain appropriate phase relations between the pulses.
In the following we will describe our theoretical approach and present convincing data from a demonstration experiment with relevance to applications in quantum information technology, i.e., DD of atomic coherences for coherent optical data storage in a \pryso crystal (termed Pr:YSO). As our numerical simulations and the experimental data show, the UR sequences outperform the best robust DD sequences available so far.

%
\emph{\textbf{The system.}}--
We consider a system, consisting of an ensemble of noninteracting two-state systems in a dephasing environment, and assume we have no control of the environment.
Similarly to previous work on robust DD sequences \cite{Alvarez13PRA}, we use a semiclassical approximation, where the free evolution Hamiltonian of a qubit includes an effective time-dependent Hamiltonian due to the system-environment interaction \cite{NMR_literature,Shore_literature}.
This is the case, e.g., when the changes in the environment are slow, compared to the delay between the pulses in the DD sequences \cite{Alvarez13PRA}.
Such systems are encountered in many solid-state spin systems, e.g., doped solids, electron spins in diamond,  electron spins in quantum dots, etc.

We denote a qubit transition frequency as
$\omega_{S}^{(k)}(t)=\overline{\omega}_{S}+\Delta^{(k)}+\epsilon(t),$
where $\overline{\omega}_{S}$ is the center frequency of the ensemble, and $\Delta^{(k)}$ is the detuning of $k$th qubit due to a slowly changing qubit-environment interaction, e.g., inhomogeneous broadening; $\epsilon(t)$ is a stochastic term due to a fast qubit-environment interaction, which cannot be refocused by DD and its effect is simply an additional exponential decay of the coherence -- we omit it further on. The assumption for constant $\Delta^{(k)}$ during a DD sequence becomes feasible by shortening the time between the pulses. This was the inspiration for the introduction of the widely used CPMG sequence, following the seminal work of Hahn \cite{Hahn50PR,CPMG_papers}. Previous experiments have demonstrated that these are reasonable assumptions for comparison of robust DD sequences \cite{Alvarez13PRA}.

\begin{figure}
\includegraphics[width=0.85\columnwidth]{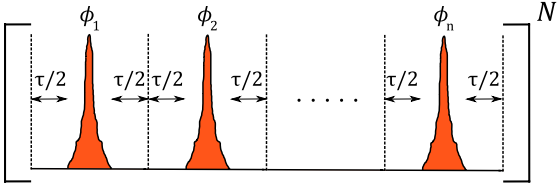}
\caption{Schematic description of a DD sequence with $n$ equally separated phased pulses. A single cycle free evolution-pulse-free evolution lies within the dashed lines. The proper choice of the relative phases of the pulses compensates both pulse errors and dephasing due to the environment. The DD sequence is repeated $N$ times during the storage time; the pulse shape can be arbitrary.
}
\label{Fig:fig1}
\end{figure}
The Hamiltonian of the system in a rotating frame at an angular frequency $\overline{\omega}_{S}$ is $\widehat{H}_{f}(t)=\Delta^{(k)}\widehat{S}_{z}$ ($\widehat{S}_{z}=\hbar\sigma_{z}/2$, $\hbar=1$) during free evolution and $\widehat{H}(t)=\widehat{H}_{f}(t)+\widehat{H}_{p}(t)$ during a pulse. The latter depends on
qubit frequency offset $\Delta^{(k)}$, the (time-dependent) detuning of the applied field $\Delta^{(p)}(t)\equiv\overline{\omega}_{S}-\omega^{(p)}(t)$ and the Rabi frequency $\Omega(t) =-\mathbf{d}\cdot\mathbf{E}(t)/\hbar$.
We make \emph{no assumptions} about $\Omega(t)$ and $\Delta^{(p)}(t)$, which may vary for the different qubits.

The dynamics of a qubit due to a pulse is described by a propagator $\textbf{U}_{\text{pulse}}$, which connects the density matrices of the system at the initial and final times $t_{\i}$ and $t_{\f}$:
 $\mathbf{\rho}(t_{\f})=\mathbf{U}_{\text{pulse}} \mathbf{\rho}(t_{\i})\mathbf{U}_{\text{pulse}}^{\dagger}$ and can be parametrized \cite{Genov2014} as
\begin{equation} \label{2stateU}
\U_{\text{pulse}}(\alpha,\beta,p) = \left[\begin{array}{cc} \sqrt{1-p}\, \e^{\i\alpha}  &  \sqrt{p}\, \e^{\i\beta} \\  -\sqrt{p}\, \e^{-\i\beta} & \sqrt{1-p}\, \e^{-\i\alpha} \end{array} \right],
\end{equation}
where $p$ is the transition probability, induced by a pulse; $\alpha$, $\beta$ are (unknown) phases.
A phase shift $\phi$ in the Rabi frequency, $\Omega(t)\to\Omega(t)\e^{\i\phi}$, is imprinted in $\U_{\text{pulse}}$ as $\beta\to\beta+\phi$
 \cite{Torosov11PRA,Torosov11PRL,Genov2014}.
The phase $\phi$ is assumed the same for every qubit (unlike $\beta$), which is usually experimentally feasible.

DD sequences traditionally consist of time-separated pulses \cite{Hahn50PR,CPMG_papers}. We consider DD with equal pulse separation (see Fig. \ref{Fig:fig1}), which was shown to be preferable for most types of environment \cite{Alvarez11PRA}. The propagator for a single cycle, defined as free evolution -- (phased) pulsed excitation -- free evolution, is $\mathbf{U}(\phi)=\U_{\text{pulse}}(\alpha+\delta,\beta+\phi,p)$, where $\delta\equiv-\Delta^{(k)}\tau/2$ accounts for the effect of the environment during free evolution.
The parameters $\alpha$, $\beta$, $\delta$, $p$ may vary for the different qubits and are affected by many factors, e.g., field inhomogeneities, effect of the environment.
The propagator of a DD sequence of $n$ free evolution-pulse-free evolution cycles, where the $k$th pulse is phase shifted by $\phi_{k}$ (see Fig. \ref{Fig:fig1}), takes the form $\mathbf{U}^{(n)}=\mathbf{U}(\phi_n)\dots\mathbf{U}(\phi_2)\mathbf{U}(\phi_1)$,
where $\phi_1,\dots,\phi_{\text{n}}$ are free control parameters. The DD sequence can be repeated $N$ times for decoupling during the whole storage time.

%
\emph{\textbf{Derivation of the UR DD sequences.}}
Our goal is to preserve an arbitrary qubit state,
which can be achieved (up to a phase shift) if a DD sequence has an even number of (phased) pulses, and each performs complete population inversion, i.e., $p=1$ (see \emph{Supplemental Material} at \cite{Supl_material}). Thus, we define our target propagator as $\U_{0}=\mathbf{U}^{(n)}(p=1)$.
We choose the phases $\phi_1,\dots,\phi_{\text{n}}$, so that systematic errors in a pulse cycle are compensated by the other cycles in a DD sequence, similarly to the technique of composite pulses \cite{Levitt84NMR}. The DD sequence performance is characterized with the fidelity \cite{RDD_review12Suter}
\begin{equation}
F=\frac{1}{2}|\text{Tr}(\U_{0}^\dagger \mathbf{U}^{(n)})|\equiv 1-\varepsilon_{\text{n}},
\end{equation}
where $\varepsilon_{\text{n}}$ is the fidelity error of a DD sequence of $n$ cycles.

In order to minimize $\varepsilon_{\text{n}}$, we perform a Taylor expansion with respect to the transition probability $p$ at $p=1$ (ideal $\pi$ pulse) and use the control parameters $\phi_{k}$ to nullify the series coefficients for every $\alpha$, $\delta$, and $\beta$ up to the largest possible order of $p$.
The phase $\phi_1$ has a physical meaning only with respect to the (unknown) phase of the initial coherence, so we take $\phi_1=0$ without loss of generality. In the case of two cycles, e.g., the well-known CPMG sequence, the fidelity error is $\varepsilon_{2}=2(1-p)\cos^2{(\alpha+\delta-\phi_2/2)}$. Thus, error compensation is not possible by a proper choice of $\phi_2$, except for a particular $\alpha+\delta$ or for certain initial states \cite{RDD_review12Suter}. However, error compensation for an arbitrary initial state becomes possible with four or a higher even number of cycles.

\begin{table}[t]
\caption{Phases of the symmetric universal rephasing (UR) DD sequences with $n$ cycles (indicated by the number in the label), based on Eq. \eqref{def:phases}.
Each phase is defined modulo $2\pi$.
} 
\begin{tabular}{l l l l} 
\hline 
Sequence & Phases & $~~\Phi^{(n)}$ \\ 
\hline 
UR4 & $(0,1,1,0)\pi$ &  $~~~\pi$ \\
UR6 & $\pm(0,2,0,0,2,0)\pi/3$ & $~\pm2\pi/3$ \\
UR8 & $\pm(0,1,3,2,2,3,1,0)\pi/2$ & $~\pm\pi/2$\\
UR10 & $\pm(0,4,2,4,0,0,4,2,4,0)\pi/5$ & $~\pm4\pi/5$ \\
UR12 & $\pm(0,1,3,0,4,3,3,4,0,3,1,0)\pi/3$ & $~\pm\pi/3$\\
UR14 & $\pm(0,6,4,8,4,6,0,0,6,4,8,4,6,0)\pi/7$ & $~\pm6\pi/7$\\
UR16 & $\pm(0,1,3,6,2,7,5,4,4,5,7,2,6,3,1,0)\pi/4$ & $~\pm\pi/4$\\
\hline 
\end{tabular}
\label{table_phases} 
\end{table}

We derive a general formula for the phases of a UR sequence of $n$ pulses (see also \emph{Supplemental Material} at \cite{Supl_material})
\begin{subequations}\label{def:phases}
\begin{align}
&\phi_k^{(n)}=\frac{(k-1)(k-2)}{2}\Phi^{(n)}+(k-1)\phi_2,\\
&\Phi^{(4m)}=\pm\frac{\pi}{m},~\Phi^{(4m+2)}=\pm\frac{2m\pi}{2m+1}.
\end{align}
\end{subequations}
The addition of an arbitrary phase $\widetilde{\phi}$ to all phases does not affect the overall performance, while $\phi_2$ can be chosen at will to perform an arbitrarily accurate phase gate
$\exp{(\i\chi\widehat{S}_{z})}, \chi=n(\phi_2-\widetilde{\phi})/2$
without additional pulses. We note that for $n=4, \phi_2=\pi/2$ we obtain the well-known XY4 sequence \cite{RDD_review12Suter}. The simplest symmetric UR DD sequences with a target $\mathbf{U}_{0}=(-1)^{n/2}\mathbf{I}$ are given in Table \ref{table_phases}.  It is notable that the order of error compensation increases \emph{linearly} with the number of cycles $n$:
\begin{equation}
\varepsilon_{n}=2(1-p)^{n/2}\sin^{2}\left[\frac{n}{2}(\alpha+\delta-\pi/2-\phi_2/2)\right].
\end{equation}
This is the central result of this Letter. Arbitrarily accurate error compensation is achievable even for small single pulse transition probability for any pulse shape, e.g., also for chirped pulses \cite{Schraft13PRA}; the linear rise in the number of pulses for higher order error compensation is superior to traditional techniques, e.g., nesting of sequences \cite{RDD_review12Suter}; the analytic formula for UR DD allows for fine tuning to the specific pulse errors and environment.

\begin{figure}
\includegraphics[width=0.96\columnwidth]{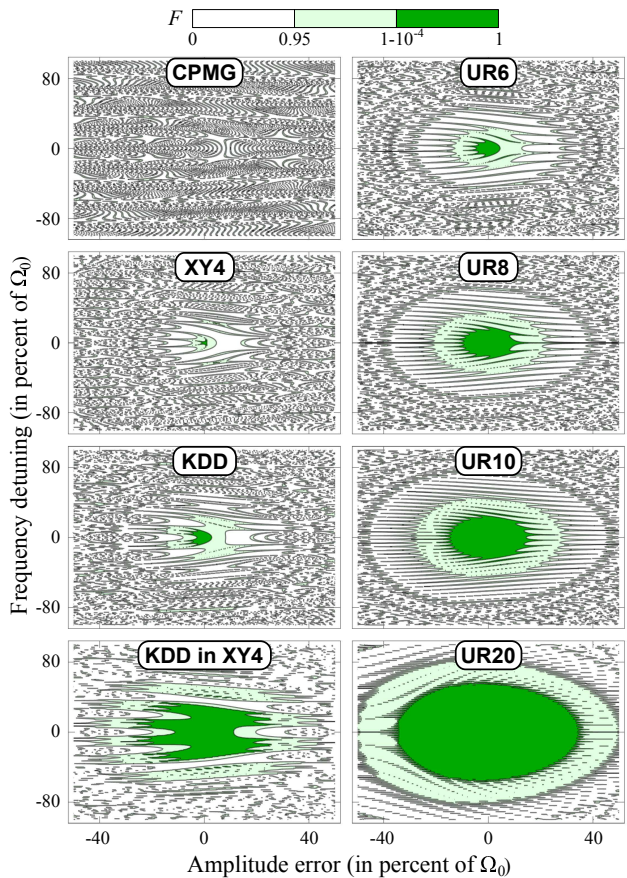}
\caption{Numerically simulated fidelity $F$ vs detuning and amplitude errors for DD sequences from Table \ref{table_phases} and \cite{RDD_review12Suter} for a total of 120 cycles. The DD pulses are rectangular with a duration of $T=\pi/\Omega_{0}$ and time separation of $\tau=4 T$.
 }
\label{Fig:fig2}
\end{figure}

Figure \ref{Fig:fig2} demonstrates the theoretical fidelity of several DD sequences
against frequency detuning and Rabi frequency errors for a single qubit. The applied rectangular pulses differ only in their phases, and each sequence is repeated to ensure a total of 120 pulses (e.g., UR10 is repeated $N=12$ times).
The parameter range corresponds to the experimental Fig. 3. The simulations show that the fidelity of the CPMG sequence is very sensitive to pulse errors, while the robustness of the UR sequences increases quickly with the sequence order. It is remarkable that the fidelity error $\varepsilon_{\text{n}}$ for UR20 stays below the $10^{-4}$ quantum information benchmark even with amplitude errors and frequency offset of nearly 40\% of the Rabi frequency. We note that the ultrahigh fidelity range expands even more with shorter pulse separation and higher order sequences. Finally, UR20 is more robust than the current state-of-the-art sequence for pulse error compensation KDD in XY4 (also of 20 pulses) \cite{RDD_review12Suter}. This is not surprising since the fidelity error $\varepsilon_{20}\sim(1-p)^{6}$ for KDD in XY4 is larger than $\varepsilon_{20}\sim(1-p)^{10}$ for UR20 ($p\to 1$).

\begin{figure}
\includegraphics[width=0.95\columnwidth]{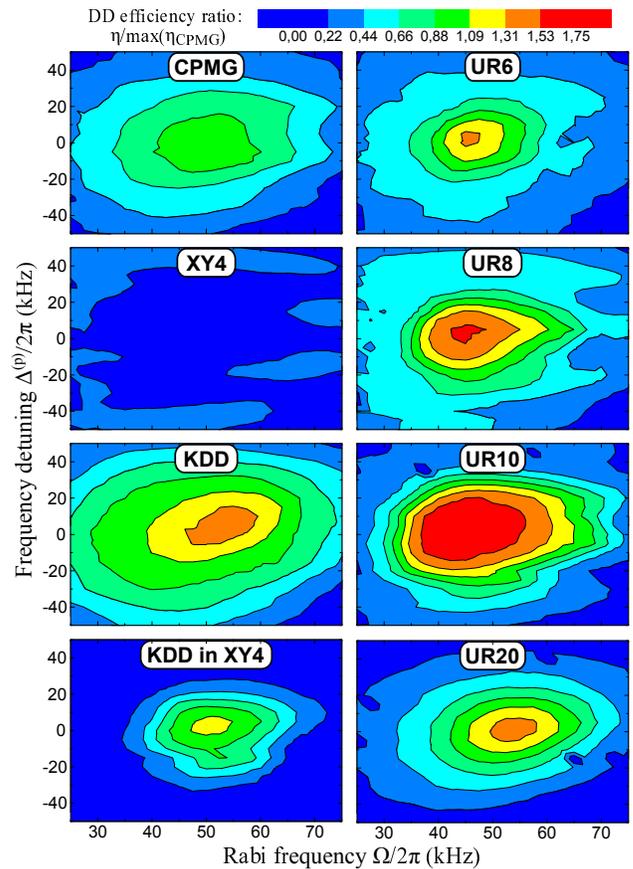}
\caption{Experimentally measured ratio of light storage efficiency for DD sequences from Table \ref{table_phases} and \cite{RDD_review12Suter}, and the maximum efficiency of the CPMG sequence. A total of $120$ rectangular pulses with duration $T=10\,\mu s$ and time separation $\tau=40\,\mu s$ were applied; storage time is 6$\,$ms. The performance of longer DD sequences is expectedly reduced by decoherence as the sequence duration approaches $T_{2}=500\,\mu$s.
}
\label{Fig:fig3}
\end{figure}

%
\emph{\textbf{Experimental demonstration.}}--
We experimentally verified the performance of the UR sequences for DD of atomic coherences for optical data storage.
In the experiment, we generate a coherence on a radio-frequency (rf) transition between two inhomogeneously broadened hyperfine levels of the Pr:YSO crystal.
The coherence
is prepared and read-out by electromagnetically induced transparency (EIT) \cite{Fleischhauer05RMP}. EIT in a doped solid was already applied to drive an optical memory with long storage times \cite{Heinze13PRL} or high storage efficiency \cite{Schraft16PRL}.
The concept and experimental setup for (single-pass) EIT light storage are described in \emph{Supplemental Material} at \cite{Supl_material} and \cite{Schraft16PRL}.

In such a coherent optical memory, it is crucial to reverse the effect of dephasing of atomic coherences during storage due to inhomogeneous broadening of the hyperfine levels ($T_{\text{deph}}\approx 13\,\mu$s). Additionally, stochastic magnetic interactions between the dopant ions and the host matrix lead to a decoherence time of $T_{2}\approx 500\,\mu s$.
DD is ideally implemented with instantaneous resonant $\pi$ pulses, which are not feasible in our experiment due to
inhomogeneous broadening and the spatial inhomogeneity of the rf field.
In order to permit a much broader operation bandwidth,
we replace the identical pulses in the widely used CPMG sequence \cite{CPMG_papers} with phased pulses. In all experiments the optical ``write'' and ``read'' sequences were kept the same, while the DD sequences with the same pulse separation have identical duty cycle (total irradiation time divided by total time) for a fair comparison; therefore, the energy of the retrieved signal measures the DD efficiency (see also \emph{Supplemental Material} at \cite{Supl_material}).

\begin{figure}
\includegraphics[width=0.96\columnwidth]{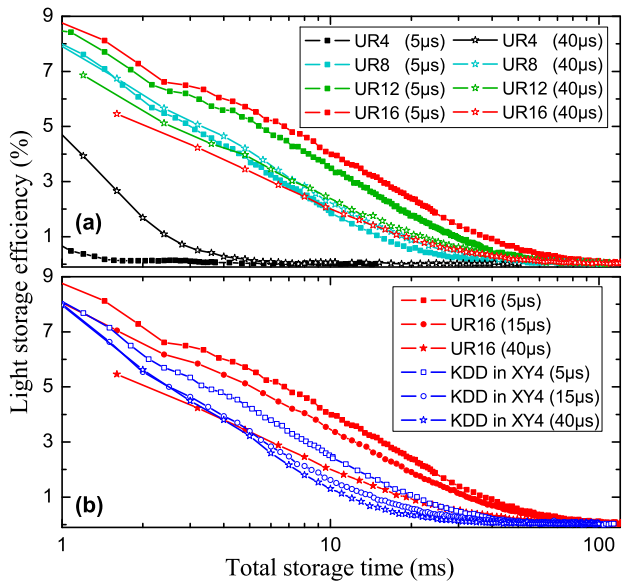}
\caption{
Experimentally measured efficiency of stored light for several DD sequences, defined in Table \ref{table_phases} and \cite{RDD_review12Suter} for different pulse separation. DD is performed with rectangular rf pulses with a frequency of $10.2$ MHz, duration of $10\,\mu$s, and a Rabi frequency $\Omega_0\approx 2\pi\,50$ kHz, optimized for a maximum efficiency with the CPMG sequence for a storage time of $100\,\mu s$. Note the logarithmic scale on the time axis.
}
\label{Fig:fig4}
\end{figure}

In the first experiment (Fig. \ref{Fig:fig3}), we compare the efficiency of several DD sequences for a storage time of 6$\,$ms, i.e., much longer than $T_{2}\approx 500\,\mu s$. Matching to the simulations in Fig. \ref{Fig:fig2}, we implement DD with 120 rectangular rf pulses.
We vary the Rabi frequency and the detuning to obtain a 2D plot of the relative storage efficiency. The experimental results confirm the theoretical prediction that the efficiency increases with the UR order until the longer sequences are significantly affected by decoherence. We note that Figs. \ref{Fig:fig2} and \ref{Fig:fig3} are expected to differ as the former simulates only the DD fidelity for a single qubit in a constant environment. For example, the CPMG sequence has a higher storage efficiency than XY4 in the experiment because it works very well for some initial quantum states \cite{RDD_review12Suter}, i.e., some atoms in the ensemble. However, applying the CPMG sequence with pulse errors effectively projects the qubits on such states, thus making it unsuitable for quantum storage. The UR sequences significantly outperform the
traditional sequences, e.g., the efficiency of UR10 is about 75\% higher than CPMG.
We also verified the superior performance of the UR sequences for a pulse separation $\tau=15\,\mu s$ and with Gaussian pulses (see \emph{Supplemental Material} at \cite{Supl_material}).

\begin{figure}
\includegraphics[width=0.99\columnwidth]{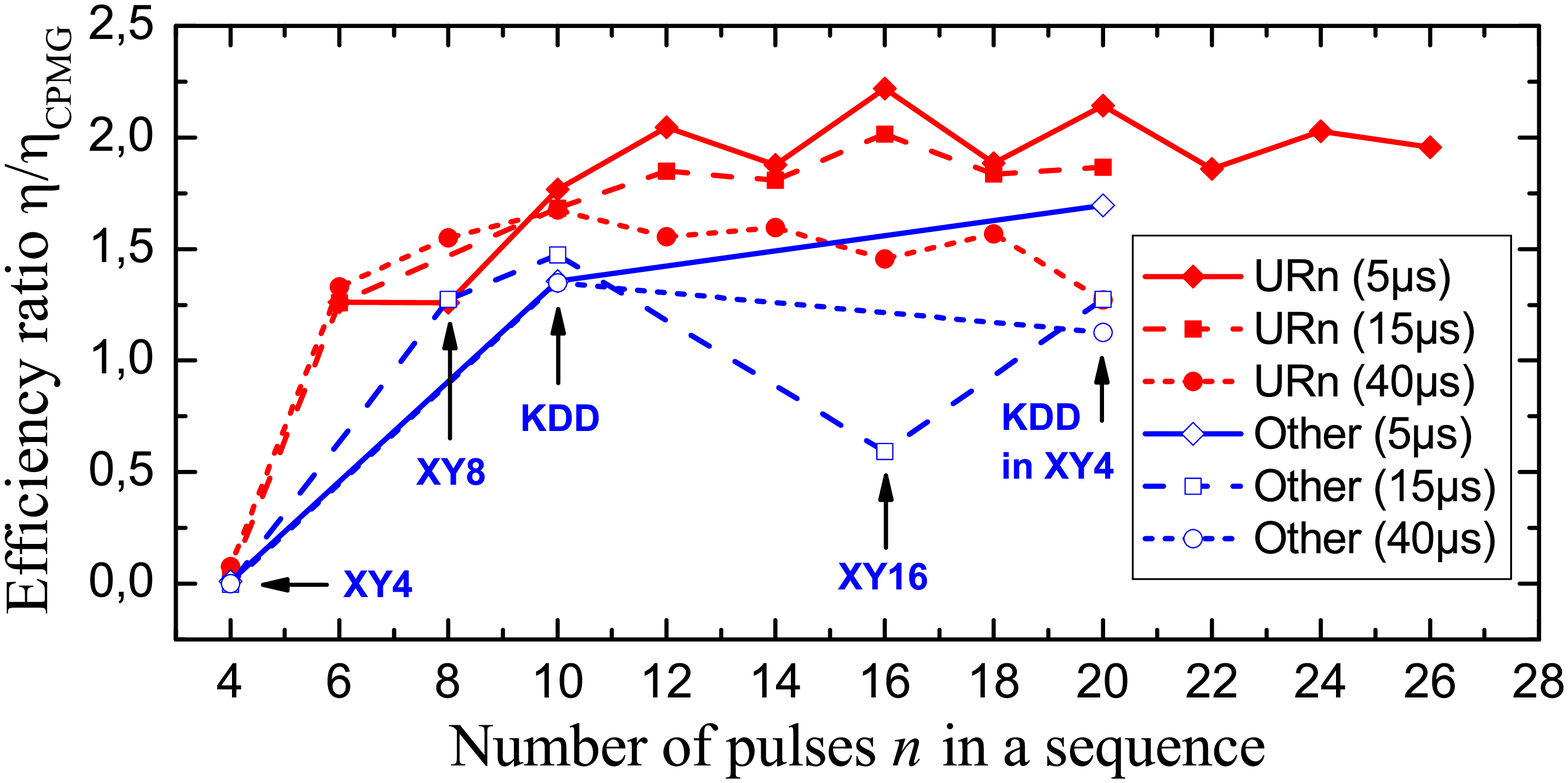}
\caption{Experimentally measured ratio of light storage efficiency for DD sequences from Table \ref{table_phases} and \cite{RDD_review12Suter}. Experimental parameters are identical to Fig. \ref{Fig:fig4}, storage time is (approximately) $6\,$ms.
}
\label{Fig:fig5}
\end{figure}

In a second experiment we compare DD efficiencies at different storage times.
Figure \ref{Fig:fig4}(a) also confirms the theoretical prediction
that the UR efficiency increases with the sequence order. The highest efficiency is achieved with UR16 for a pulse separation of $\tau=5\,\mu s$, while UR12 is better for $40\,\mu s$.
This is explained by the trade-off between longer sequences that compensate pulse errors better and shorter sequences that suffer less from decoherence.
The UR analytical formula and the fast improvement in error compensation allow for fine-tuning of the optimal sequence to the specific environment.
Shorter (than $5\,\mu s$) pulse separation and even continuous UR sequences are theoretically possible and should provide even better performance. However, these were not possible in our experimental setup.
Figure \ref{Fig:fig4}(b) compares the performance of the UR16 and the state-of-the-art KDD in XY4 sequence \cite{RDD_review12Suter}. The experimental data show that UR16 performs remarkably better than KDD in XY4.
This is expected from theory since UR16 has both higher order pulse error compensation and less pulses.
The improvement is less for $\tau=40\,\mu s$ as the duration of both sequences exceeds $T_2$.

Figure \ref{Fig:fig5} summarizes the experimental performance of UR and other sequences \cite{RDD_review12Suter} vs. number of pulses $n$ in a sequence. The data confirm the UR superior performance and the improvement of error compensation with $n$. The optimal sequence changes with pulse separation due to the trade-off between pulse errors and decoherence during a sequence, which affects its error self-compensatory mechanism \cite{RDD_review12Suter}.
The slight oscillation in UR efficiency is likely due to higher-order effects for the particular time separation.
Finally, we note that we also verified the superior performance of the UR sequences in comparison to composite pulses with the same duty cycle, e.g., U5a and U5b \cite{Genov2014}.

%
\emph{\textbf{Conclusion.}}--
We theoretically developed and experimentally demonstrated universally robust DD sequences, which compensate systematic errors in any experimental parameter and the effect of a slowly changing dephasing environment to an arbitrary order for any pulse shape and initial condition.
The UR sequences require a linear growth in the number of pulses for higher order error compensation, which is faster than traditional methods, e.g., nesting of sequences. The only assumptions made are those of a coherent evolution during the DD sequence (the correlation time of the environment is longer than the sequence duration ) and identical phased pulses. We also experimentally confirmed the superior performance of our UR sequences for DD
for coherent optical data storage in a Pr:YSO crystal.

%
\begin{acknowledgments}
We acknowledge experimental support by M. Hain and J. Zessin (Technische Universit{\"a}t Darmstadt). This work is supported by the Deutsche Forschungsgemeinschaft and the Alexander von Humboldt Foundation.
\end{acknowledgments}


\end{document}


\author{Genko T. Genov}
\affiliation{Institut f{\"u}r Angewandte Physik, Technische Universit{\"a}t Darmstadt, Hochschulstr. 6, 64289 Darmstadt, Germany}
\author{Daniel Schraft}
\affiliation{Institut f{\"u}r Angewandte Physik, Technische Universit{\"a}t Darmstadt, Hochschulstr. 6, 64289 Darmstadt, Germany}
\author{Nikolay V. Vitanov}
\affiliation{Department of Physics, St. Kliment Ohridski University of Sofia, 5 James Bourchier blvd, 1164 Sofia, Bulgaria}
\author{Thomas Halfmann}
\affiliation{Institut f{\"u}r Angewandte Physik, Technische Universit{\"a}t Darmstadt, Hochschulstr. 6, 64289 Darmstadt, Germany}

\title{Supplemental Material to Arbitrarily Accurate Pulse Aequences for Robust Dynamical Decoupling}

\date{\today}

\maketitle

\emph{\textbf{Detailed derivation of the arbitrarily accurate UR DD sequences.}}
Our goal is to preserve an arbitrary quantum state of a two-state system by reducing the average effect of its unwanted interaction with a dephasing environment by DD. Additionally, we aim at simultaneously minimizing the effect of pulse errors of our imperfect rephasing pulses. In the ideal case, this implies that $\mathbf{U}^{(n)}$ is equal to the identity propagator or the propagator of a target phase gate $\exp{(\i \chi \widehat{S}_{z})}$.

We note two important observations: (1) the transition probability of a pulse cycle (free evolution - pulse - free evolution) is the same as the transition probability of the applied rephasing pulse, (2) complete single pulse population inversion, i.e., $p=1$, is a sufficient condition for nullifying the detrimental effect of an (assumed constant) dephasing system-environment interaction if we apply an even number of DD cycles because (see main text for a definition of $\mathbf{U}^{(n)}$)
\begin{equation}\label{eq:U0}
\mathbf{U}^{(n)}(p=1)=(-1)^{n/2}\exp{\left(\i\chi \widehat{S}_{z}\right)},~n=2m
\end{equation}
for arbitrary $\alpha$, $\beta$, and $\delta$, where $\chi=2\sum_{k=1}^{n/2}\left(\phi_{2k}-\phi_{2k-1}\right)$. As we can see, the propagator $\mathbf{U}^{(n)}(p=1)$ does not depend on $\alpha$, $\beta$, and $\delta$, but only on the relative phases between the pulses $\phi_{k}$, which we assume to control.
Thus, $\mathbf{U}^{(n)}(p=1)$ does not depend on the effect of the environment and we can choose the phases $\phi_{k}$ to make it the identity propagator $\mathbf{I}$ or the propagator of a target phase gate $\exp{(\i \chi \widehat{S}_{z})}$.
Therefore, we apply an even number of cycles in a DD sequence in the further analysis, and we define our target propagator $\mathbf{U}_{0}=\mathbf{U}^{(n)}(p=1)$ as given in Eq. \eqref{eq:U0}.

Pulses in actual experiments are often far from perfect, e.g., due to limited bandwidth, field inhomogeneity, and the (assumed constant) unwanted effect of the environment during a DD sequence. We require our DD sequences to work for any arbitrary quantum state, and their performance is traditionally measured with the fidelity \cite{RDD_review12Suter}
\begin{equation}
F=\frac{|\text{Tr}(\U_{0}^\dagger \mathbf{U}^{(n)})|}{\sqrt{\text{Tr}(\U_{0}^\dagger \mathbf{U}_{0})\text{Tr}(\mathbf{U}^{(n)\dagger} \mathbf{U}^{(n)})}}=\frac{1}{2}|\text{Tr}(\U_{0}^\dagger \mathbf{U}^{(n)})|,
\end{equation}
where $\mathbf{U}_{0}$ is the target propagator, $\mathbf{U}^{(n)}$ is the propagator of the DD sequence.
We define $\varepsilon_{\text{n}}\equiv 1-F$ as the fidelity error of a DD sequence of $n$ cycles.

In order to minimize $\varepsilon_{\text{n}}$, we perform a Taylor expansion with respect to the transition probability $p$ at $p=1$ (ideal $\pi$ pulse) and use the control parameters $\phi_{k}$ to nu	llify the series coefficients for every $\alpha$, $\delta$, and $\beta$ up to the largest possible order of $p$.
We note that the phase $\phi_1$ of the first applied pulse has a physical meaning only with respect to the phase of the initial coherence, which is often unknown. Therefore, we take $\phi_1=0$ without loss of generality. In the case of a DD sequence of two cycles, i.e., the well-known CPMG sequence, the fidelity error $\varepsilon_{2}=2(1-p)\cos^2{(\alpha+\delta-\phi_2/2)}$, i.e., higher-order error compensation is not possible by a proper choice of $\phi_2$, except for a particular $\alpha+\delta$ or for certain initial states \cite{RDD_review12Suter}. However, error compensation for an arbitrary initial state becomes possible by a proper choice of the phases of the applied arbitrary pulses with four or a higher even number of cycles. For example, in the case of a DD sequence of four pulses, we obtain
\begin{subequations}\label{eq:phases_UR4}
\begin{gather}
\phi_{3}=2\phi_2+\pi,~\phi_{4}=3\phi_2+\pi,\\
\varepsilon_{4}=2(1-p)^2\sin^2{(2\alpha+2\delta-\phi_2)},~\chi=4\phi_2,
\end{gather}
\end{subequations}
where $\phi_2$ can be chosen appropriately to obtain a dynamically corrected phase gate. For example, taking $\phi_2=0,\pm\pi/2,\pi$ allows us to obtain a DD sequence with an identity propagator up to a global phase $\mathbf{U}_{0}=\pm\mathbf{I}$ ($\chi=0$) with a fidelity error of the order of $(1-p)^2$ for arbitrary $\alpha$, $\beta$, and $\delta$. Since the fidelity error $\varepsilon_{4}\sim O(1-p)^2<O(1-p)$ of standard CPMG (usually $p\to 1$), we achieve improved robustness with the UR4 sequences, defined in Eq. \eqref{eq:phases_UR4}, in comparison to CPMG. We note that the improvement is achieved without any assumptions for the initial condition, pulse shape, pulse errors and the effect of the (assumed constant) dephasing environment during the DD sequence. We also note that for $\phi_2=\pi/2$ we obtain the well-known XY4 sequence \cite{RDD_review12Suter}. An example of a simple symmetric UR4 sequence is given in Table I in the main manuscript for the the case when $\phi_2=\pi$.

We then derive a general formula for the phases of a UR sequence of $n$ pulses (see also the main manuscript)
\begin{subequations}\label{def:phases}
\begin{align}
&\phi_k^{(n)}=\frac{(k-1)(k-2)}{2}\Phi^{(n)}+(k-1)\phi_2,\\
&\Phi^{(4m)}=\pm\frac{\pi}{m},~\Phi^{(4m+2)}=\pm\frac{2m\pi}{2m+1}.
\end{align}
\end{subequations}
The addition of an arbitrary phase $\widetilde{\phi}$ to all phases does not affect the overall performance, while $\phi_2$ can be chosen at will to perform an arbitrarily accurate phase gate
$\exp{(\i\chi\widehat{S}_{z})}, \chi=n(\phi_2-\widetilde{\phi})$.

\begin{figure}[t]
\includegraphics[width=0.45\columnwidth]{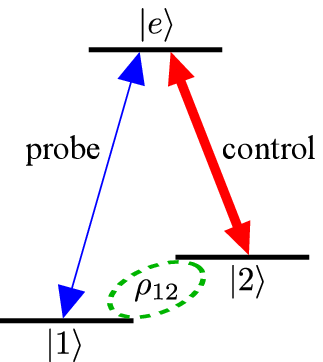}
\caption{(Color online)
Level scheme for EIT light storage.
}
\label{Fig:fig1EITsm}
\end{figure}

\emph{\textbf{EIT light storage efficiency and DD performance.}}
Light storage by electromagnetically-induced transparency (EIT) is usually implemented in a $\Lambda$-type atomic medium (see Fig. \ref{Fig:fig1EITsm} for an idealized scheme), where the atoms are initially prepared with all population in the ground state $|1\rangle$ \cite{Fleischhauer05RMP,Novikova12review}. EIT uses a strong control pulse, tuned to the transition between the ground state $|2\rangle$ and an excited state $|e\rangle$. The coherent interaction with the control pulse modifies the optical properties of the system. Specifically, the medium becomes transparent for a probe pulse on the $|1\rangle\leftrightarrow|e\rangle$ transition.
Moreover, the group velocity of the probe pulse is significantly reduced due to the strong control field and the probe pulse is, thus, compressed in the storage medium.

By reducing the control pulse intensity adiabatically, the ``slow light'' probe pulse is ``stopped'' and converted into an atomic coherence $\rho_{12}$ of the quantum states $|1\rangle$ and $|2\rangle$ along the probe propagation path $z$ in the atomic medium. This establishes a spin wave of spatially distributed atomic coherences $\rho_{12}(z,t)$ in the medium, which contains all information of the incoming probe pulse \cite{Novikova12review}. This is usually termed the \emph{``write''} process of EIT light storage and in the perfect case it maps \cite{Fleischhauer05RMP}
\begin{equation}
\mathcal{E}_{\text{in}}(z,t)\to\sqrt{N/V}\rho_{12}(z,t),
\end{equation}
where $\mathcal{E}_{\text{in}}(z,t)$ is the electric field envelope of the probe pulse, $N/V$ is the number density of atoms, $\rho_{12}(z,t)$ is the coherence at position $z$ in the ensemble.

During \emph{``storage''}, we apply DD sequences to reduce the effect of the qubit-environment interaction that causes decoherence. This leads to
\begin{equation}
\rho(z,t+T_{\text{st}})=\mathbf{U}_{\text{DD}}(z,t+T_{\text{st}})\rho(z,t)\mathbf{U}_{\text{DD}}^{\dagger}(z,t+T_{\text{st}}),
\end{equation}
where $T_{\text{st}}$ is the storage time, and $\mathbf{U}_{\text{DD}}(z,t+T_{\text{st}})$ is a propagator that depends on the applied DD sequence, the static effect of the environment (e.g., due inhomogeneous broadening, field inhomogeneity along $z$, etc.), as well as the specific noise realization for a particular atom, e.g., due to changes in its environment. Experimentally, we are interested in the ensemble average at $z$ at time $t+T_{\text{st}}$: $\langle \rho(z,t+T_{\text{st}})\rangle$.

The spin wave can be read out after some storage time $T_{\text{st}}$ by applying a control read pulse to beat with the atomic coherences and generate a signal pulse on the $|1\rangle\leftrightarrow|e\rangle$ transition. This is usually termed the \emph{``read''} process, which in the perfect case maps back the spin wave onto an electromagnetic field on the probe  pulse transition \cite{Fleischhauer05RMP}
\begin{equation}
-\sqrt{N/V}\langle\rho_{12}(z,t+T_{\text{st}})\rangle\to \mathcal{E}_{\text{out}}(z,t+T_{\text{st}}).
\end{equation}
Light storage efficiency is determined by the ratio of the energy of retrieved photons after storage vs. the energy of the photons of the input probe pulse \cite{Fleischhauer05RMP,Novikova12review}
\begin{equation}
\eta=\frac{\int_{T_{\text{st}}}^{\infty}| \mathcal{E}_{\text{out}}(z_{\text{f}},t)|^2\d t}{\int_{-T_{\text{pr}}}^{0}| \mathcal{E}_{\text{in}}(z=0,t)|^2\d t},
\end{equation}
where $T_{\text{pr}}$ is the probe pulse duration, $z=0$ and $z_{\text{f}}$ are the beginning and end of the storage medium.

In our experiment, we keep the \emph{``write''} and \emph{``read''} procedure identical for every DD sequence. Since $\rho_{12}(z,t)$ after the \emph{``write''} process is in principle unknown and its amplitude and phase can also vary spatially, DD efficiency and light storage efficiency are maximized if  $\mathbf{U}_{\text{DD}}(z,t+T_{\text{st}})\to I$, i.e., the identity propagator. Therefore, the overall light storage efficiency provides a useful measure for evaluating the performance of DD sequences, and it has been applied before in similar experiments in optical memories \cite{Suter_DD_light_storage}.

\begin{figure}[H]
\includegraphics[width=\columnwidth]{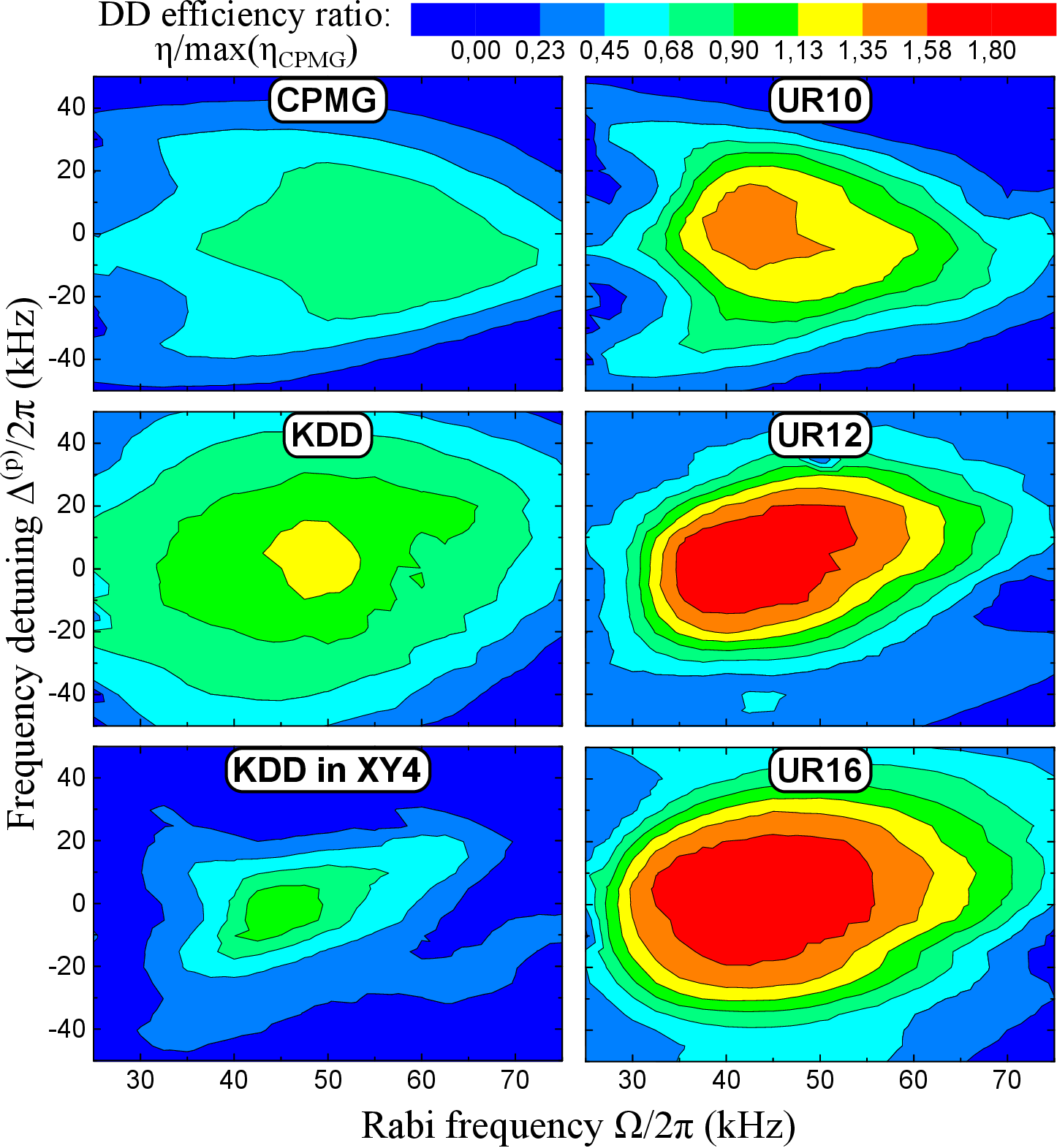}
\caption{(Color online)
Experimentally measured ratio of light storage efficiency for several DD sequences, defined in \cite{RDD_review12Suter} and Table I in the main text, and the maximum efficiency of CPMG. A total of $240$ rectangular pulses with duration $T=10\,\mu$s and time separation $\tau=15\,\mu$s were applied; storage time is 6$\,$ms.
}
\label{Fig:fig2sm}
\end{figure}

\begin{figure}[H]
\includegraphics[width=\columnwidth]{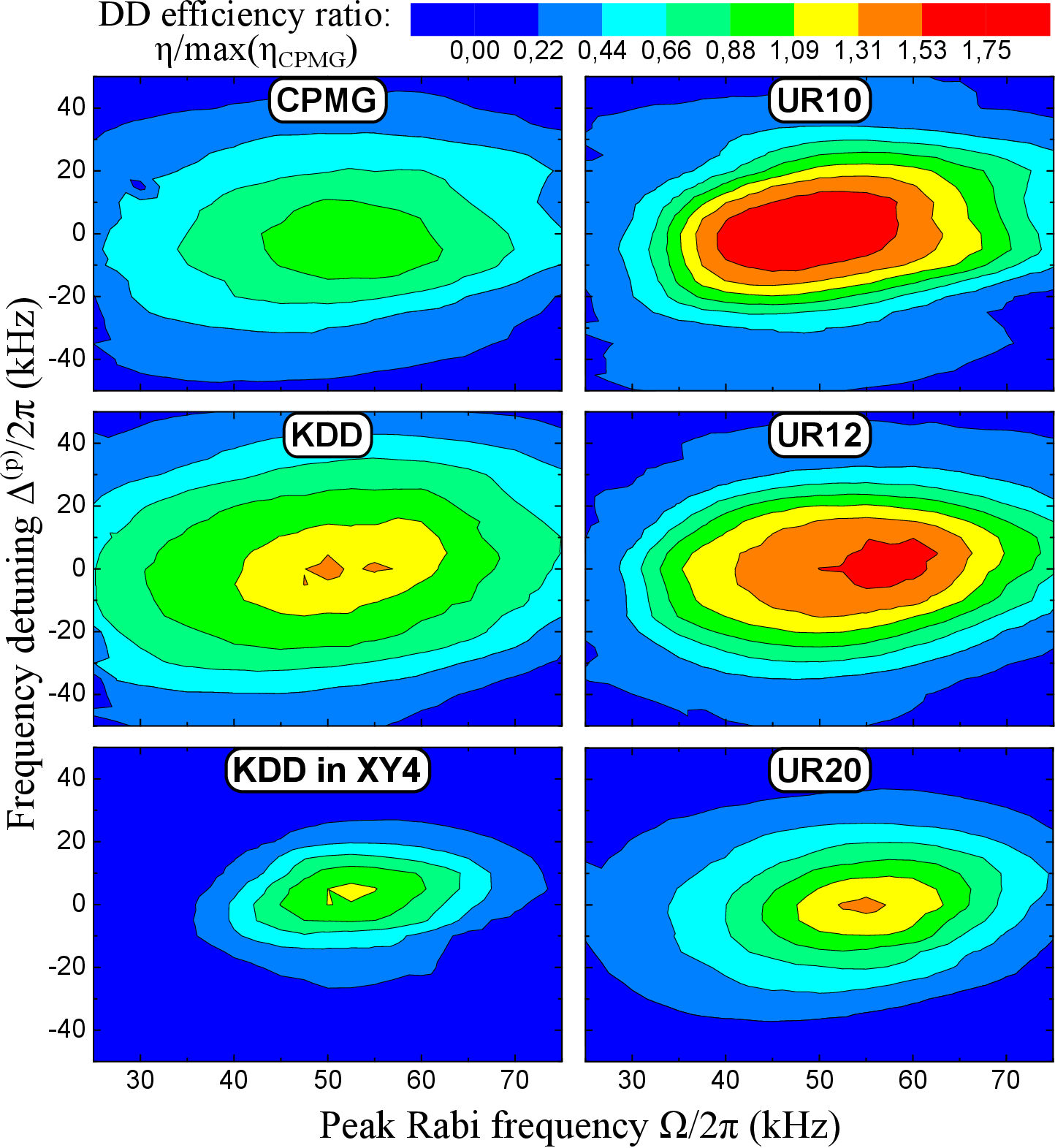}
\caption{(Color online)
Experimentally measured ratio of light storage efficiency for several DD sequences, defined in \cite{RDD_review12Suter} and Table I in the main text, and the maximum efficiency of CPMG. A total of $120$ (truncated) \emph{Gaussian} pulses with duration FWHM$=9 \mu\,$s, $T=3~\text{FWHM}$, and $\tau=23 \mu\,$s were applied; storage time is 6$\,$ms.
}
\label{Fig:fig3sm}
\end{figure}

\emph{\textbf{Additional experimental data: shorter pulse separation and Gaussian pulses.}}
Figure \ref{Fig:fig2sm} and Fig. \ref{Fig:fig3sm} show additional experimental data, which complement Fig. 3 in the main text. The phases of the UR sequences are given in Table I in the main text, while the ones of the other sequences can be found in \cite{RDD_review12Suter}.

In the first experiment (see Fig. \ref{Fig:fig2sm}) we compare the storage efficiency of several DD sequences for a storage time of 6$\,$ms, similarly to Fig. 3 in the main text. We perform DD with 240 rectangular RF pulses (compared to 120 in Fig. 3 in the main text) with a duration of $T=10\,\mu$s and a shorter pulse separation $\tau=15\,\mu$s. Again, we vary the Rabi frequency and the detuning of the applied pulses to obtain a 2D plot of the storage efficiency.

The experimental results show that, as expected from theory, the storage efficiency increases with the order of the UR sequences. The UR sequences significantly outperform the
other traditional sequences with the same number of pulses, e.g., the highest storage efficiency of UR12 and UR16 is about 80 per cent higher than CPMG. Unlike Fig. 3 in the main text where UR10 was the top performer, the best performing sequences are UR12 and UR16 due to the shorter time separation between the pulses.

Again, this is explained by the requirement that the DD sequences must be shorter than the correlation time of the environment. Thus, there is a trade-off between longer sequences that compensate pulse errors and shorter sequences that suffer less from decoherence. The shorter time separation of $\tau=15\,\mu$s is preferable to $\tau=40\,\mu$s in the main text, as both the absolute light storage efficiency and the improvement relative to CPMG are higher. Shorter pulse separation and even continuous application of the UR sequences are theoretically possible and should provide even better performance. However, these were not possible in our experimental setup.

In another experiment (see Fig. \ref{Fig:fig3sm}), we demonstrate that the UR sequences perform error compensation with pulses of a different shape, i.e., \emph{Gaussian} pulses. Again, we compare the storage efficiency of several DD sequences for a storage time of 6$\,$ms. We perform DD with 120 (truncated) \emph{Gaussian} RF pulses with a FWHM = $9\,\mu$s, duration of $T=3$ FWHM and a pulse separation $\tau=23\,\mu$s. A single free evolution - pulse - free evolution cycle duration is $T+\tau=50\,\mu$s, i.e., the same as the one of the rectangular pulses in Fig. 3 in the main text.

We vary the Rabi frequency and the detuning of the applied pulses to obtain a 2D plot of the storage efficiency. The experimental results show that, as expected from theory, the UR sequences significantly outperform the other traditional sequences with the same number of pulses. The highest storage efficiency is obtained with the UR10 sequence, similarly to the case of rectangular pulses, described in Fig. 3 in the main text. Thus, we demonstrated that the UR sequences improve the DD performance not only for rectangular pulses but also for other pulse shapes, e.g., Gaussian pulses.
